\documentstyle[seceq,preprint,epsf,mbf]{ptptex}

\newcommand{\bbone}{{\mathchoice {\rm 1\mskip-4mu l} {\rm 1\mskip-4mu l}
{\rm 1\mskip-4.5mu l} {\rm 1\mskip-5mu l}}}

\renewcommand{\theequation}{\thesection.\arabic{equation}}

\newcommand{\Hmu}{{\hat \mu}}

\newcommand{\tr}{{\rm tr}}
\newcommand{\lpm}{\lambda^{\pm}}
\newcommand{\ab}{{\alpha\beta}}
\newcommand{\gwzg}{{\Gamma_{\scriptscriptstyle WZ}[g]}}

\newcommand{\qc}{,}
\newcommand{\qp}{.}
\newcommand{\lp}{\left(}
\newcommand{\rp}{\right)}
\newcommand{\Inc}{\frac{1}{\lambda^2}}
\newcommand{\Hdel}{{\hat \partial}}
\notypesetlogo  
\preprintnumber{
KUNS-1409\\ HE(TH)~96/10 \\ hep-lat/9608137
}
\title{
Wess-Zumino term by Vacuum Overlap Formula
}
\author{
Yoshio {\sc Kikukawa}\footnote{E-mail address:
kikukawa@gauge.scphys.kyoto-u.ac.jp}   and
Shunji {\sc Miyazaki}\footnote{E-mail address:
miyazaki@gauge.scphys.kyoto-u.ac.jp}
}
\inst{
Department of Physics, Kyoto University \\ Kyoto 606-01, Japan
}

\recdate{
\today
}

\abst{
We examine the vacuum overlap formula for
the two-dimensional SU(2) Wess-Zumino term in lattice regularization.
Perturbatively it reproduces the Wess-Zumino term correctly
in the continuum limit and yields the IR fixed point in the beta
function of the chiral model.
Nonperturbatively it shows a sharp Gaussian distribution
for the SU(2) chiral field configurations
in the scaling region, where smooth 
configurations 
dominate even in the symmetric phase due to asymptotic freedom.
Crossover is sharp from the strong coupling region where
the Wess-Zumino term fluctuates hard and the species doublers'
contribution is suspected to affect it.
}

\begin{document}

\maketitle

\section{Introduction}
The vacuum overlap
formula\cite{olnn,twist,schwinger,notwave,majorana,2dtorus_bc}
gives a well-defined method to
calculate the chiral fermion determinant
in lattice regularization.
By this formula,
one first considers two Wilson fermion fields
which posses positive mass and negative mass, respectively.
They are assumed to couple to the link variables of
given background gauge field.
We next solve the ground states of the two fields.
Then the overlap of the two ground states gives
the chiral determinant.
There remains an ambiguity in the complex phase of the determinant.
It is fixed by
the convention that the inner product of each ground state
with the corresponding free ground state is real and positive.

The complex phase of the vacuum overlap
is able to reproduce the full complex phase of chiral
determinant correctly for a smooth background gauge field.
The complex phase is gauge noninvariant and its variation under a
gauge transformation was shown to give the consistent
anomaly.\cite{olnn,shamir_anomaly,al,rs}
In the continuum U(1) chiral gauge theory which is defined on the
two-dimensional torus, the chiral determinant with the uniform
background gauge potential (toron) can be calculated exactly.
The vacuum overlap reproduces this exact result.\cite{twist}

Actually, this lattice definition of the complex phase
of chiral determinant is closely related to the $\eta$ invariant
definition in the continuum theory.\cite{eta_inv}
The vacuum overlap formula can be derived from the $(2n+1)$
dimensional system of Wilson fermion with kink mass\cite{kaplan}
by taking infinite limit of the extra dimension,
provided that appropriate Pauli-Villars (Wilson) fields are
included.\cite{olnn,domainwall-ol-gs,notwave}
For this limiting procedure to be well-defined,
it also needs that the background gauge field is topologically
trivial.\cite{notwave}
By virtue of the Pauli-Villars fields,
we may take the {\it naive} continuum limit $a \rightarrow 0$
in this formula keeping the kink and Pauli-Villars masses finite.
Then it reduces to the formula which defines the $\eta$ invariant.
The only technical difference is that
the smooth interpolation function of two gauge fields is replaced
by the sharp step function at the boundary of the wave guide of
gauge interaction.

For generic gauge fields on the lattice, however,
several authors\cite{waveguide,domainwall-ol-gs,notwave,2dtorus_bc,gribov}
argued that the nature of gauge degrees of freedom may spoil the chiral
structure of the formula.
For the $(2n+1)$ dimensional Wilson fermion with the kink mass
and the wave guide of gauge interaction,\cite{kaplan}
it was shown\cite{waveguide}
that the large fluctuation of gauge freedom
induces another light fermion mode of opposite chirality at the
boundary of the wave guide.
Such a light mode could also give the contribution to the complex phase of
chiral determinant.

Singular gauge transformations can also affect the complex phase of
the vacuum overlap.
In the continuum U(1) gauge theory which is defined
on the two-dimensional torus, we can consider the singular gauge
transformation which wipe out the toron into the singular potential
with delta functions.
Then, it was shown\cite{2dtorus_bc} that the gauge invariance breaks by
the singular gauge transformation even in the anomaly free U(1) chiral
gauge theory, provided that the background toron is {\it large} and
the anti-periodic boundary condition is chosen for all fermions.
The vacuum overlap reproduces the result as well.
The authors of ref.~\cite{2dtorus_bc} claimed that it is these
singular gauge transformations which spoils the
chiral phase when averaged on the gauge orbit.
\footnote{
For the {\it small} toron,
the chiral determinant in the continuum limit is gauge invariant
even under the singular gauge transformation.\cite{2dtorus_bc}
Choosing such {\it small} toron, the gauge integration of chiral
determinant was performed.\cite{gribov}
The result shows that the large gauge fluctuation affects
the chiral phase even for the {\it small} toron and seems to
cause it to vanish. }

Therefore, it is a crucial problem in this approach to control and
reduce the fluctuation of gauge degree of freedom.
In this paper, we examine the gauge freedom of the
two-dimensional SU(2) gauge field.
In order to control and reduce the fluctuation, we introduce
the kinetic term for the variable of gauge freedom,
which consists of the nearest neighbor coupling.
We calculate the Wess-Zumino term
by the vacuum overlap formula and measure its distributions.
The whole system we consider here is the lattice counterpart of
the SU(2) Wess-Zumino-Witten model off the critical point.

This type of kinetic term for gauge freedom was examined in the
U(1) case.\cite{gauge_2dXY}
It was shown that the reduction of the fluctuation is not sufficient
near the critical point of the corresponding X-Y model.
On the contrary, in the case of SU(2) chiral field considered here,
the coupling in the critical region is weak because of the asymptotic
freedom. It is expected that the fluctuation can be reduced in this
region even thought the SU(2) chiral symmetry is realized linearly.

This paper is organized as follows. In section 2,
the lattice counterpart of the two-dimensional
SU(N) Wess-Zumino-Witten model is examined in the lattice
perturbation theory.
We show that the vacuum overlap formula
reproduces the Wess-Zumino term correctly in the continuum limit.
We also calculate the beta function of the model and show
that it has the nontrivial IR fixed point, as in the continuum theory.
In section 3, the Wess-Zumino term is examined numerically.
The SU(2) chiral field is generated by the Monte Carlo method
(Cluster algorithm\cite{cluster})
with the action of the nearest neighbor coupling
for several values of the coupling constant.
With these configurations, we calculate the complex phase of the
vacuum overlap and its distribution is obtained.
In section 4, we give a discussion concerning
the numerical estimate of observables in the Wess-Zumino-Witten model
by incorporating the imaginary Wess-Zumino action.

\newpage
\section{Two-dimensional Wess-Zumino term on the Lattice}

\subsection{Vacuum overlap formula and Wess-Zumino term}
\label{subsubsec:WZterm}
The definition of the vacuum overlap formula
for the lattice regularized chiral determinant in $2n$ dimension
is given by \cite{olnn}
\begin{equation}
\det({\bf C})_{\rm reg} \equiv \frac{ _{\bbone}\langle -|-\rangle _U
\,{_U}\langle -|+\rangle _U
\,{_U}\langle +|+\rangle _{\bbone}}{|_{\bbone}\langle -|-\rangle _U|
\,{_{\bbone}}\langle -|+\rangle _{\bbone}
\,|{_U}\langle +|+\rangle _{\bbone}|}\qp
\end{equation}
$|\pm \rangle_U$ are the vacuum state of the second quantized
Hamiltonian ${\mbf H}^{\pm}$, which defined by
\begin{eqnarray}
{\mbf H}^{\pm}(U)&=&\sum_{n\alpha i}
\sum_{m\beta j} a^{\dagger}_{n\alpha i} H^{\pm}_{n\alpha i,m\beta j}
 a^{}_{m\beta j}\qc\\
H^{\pm}(U)&=&\gamma_5[\mp m_0+C+B]\qc \label{6}\\
C_{n\alpha i,m\beta j}&=&\frac{1}{2}\sum_{\mu}(\gamma_\mu)_{\alpha\beta}
(\delta_{m,n+\Hmu}U_{n,\mu}-\delta_{n,m+\Hmu}U^{\dagger}_{m,\mu}) \qc\\
B_{n\alpha i,m\beta j}&=&\frac{r}{2}\sum_{\mu}\delta_\ab(2\delta_{n,m}
-\delta_{m,n+\Hmu}U_{n,\mu}-\delta_{n,m+\Hmu}U^{\dagger}_{m,\mu}) \qc
\end{eqnarray}
where we set the lattice spacing $a$ to unity, $m_0$ is a bare mass
parameter , $C$ and $B$ are the massless Dirac operator and the Wilson
term respectively. $a_{n\alpha i},a^{\dagger}_{n\alpha i}$
are field operators satisfying the following anti-commutation relations :
\begin{eqnarray}
\{a_{n\alpha i},a^{\dagger}_{m\beta j}\}&=&
\delta_{nm}\delta_\ab\delta_{ij}\qc\nonumber\\
\{a_{n\alpha i},a_{m\beta j}\}
&=&\{a_{n\alpha i}^\dagger,a_{m\beta j}^\dagger \}=0\qp
\end{eqnarray}
The vacuum overlaps, $ \,_U\langle -|+\rangle_U\,$ etc., can be expressed
by the one particle wave functions $\psi^{\pm}_K(n,U)$, which are
the eigenfunctions of $H^{\pm}(U)$ with positive eigenvalues labeled $K$:
\begin{eqnarray}
_U\langle- |+\rangle_U=\mathop{\det}_{\scriptscriptstyle K,K'}
 \lp \sum_n\psi^-_K(n,U)^{\dagger}\psi^+_{K'}(n,U)\rp\qc\\
_U\langle\pm |\pm\rangle_{\bbone}
=\mathop{\det}_{\scriptscriptstyle K,K'}
 \lp \sum_n\psi^{\pm}_K(n,U)^{\dagger}\psi^{\pm}_{K'}(n,\bbone)\rp\qc
\end{eqnarray}
where color and spin indices are suppressed.

The gauge transformation of the link variables is defined by
\begin{equation}
U_{n,\mu}\rightarrow U^g_{n,\mu}=g_n^\dagger U_{n,\mu}g_{n+\Hmu}\qp
\label{1}
\end{equation}
where $g_n$ are elements of gauge group SU($N$) and $U^g$ denote
the gauge transformation of $U$.

By this gauge transformation, $\psi^{\pm}_K(n,U)$ transform as
\begin{equation}
\psi^{\pm}_K(n,U)\rightarrow\psi^{\pm}_K(n,U^g)=
g_n^{\dagger}\psi^{\pm}_K(n,U)\qc
\end{equation}
Then the vacuum overlap $_U\langle- |+\rangle_U$ is gauge invariant
but the others are gauge dependent and are expressed as
\begin{equation}
_{U^g}\langle\pm |\pm\rangle_{\bbone}=\mathop{\det}_{\scriptscriptstyle K,K'}
 \lp\sum_n\psi^{\pm}_K(n,U)^{\dagger}g_n\psi^{\pm}_{K'}(n,\bbone)\rp\qp
\label{3}
\end{equation}

The variation of the complex phase of the chiral determinant by gauge
transformation gives the gauged Wess-Zumino
effective action.\cite{wz}
The corresponding quantity of the vacuum overlap formula
is written by
\begin{eqnarray}
{\cal W}^{\pm}(U,g)&\equiv&\frac{ _{U^g}\langle \pm|\pm\rangle _{\bbone}}
{ |{_{U^g}}\langle \pm|\pm\rangle _{\bbone}|}\qc\\
\exp(i\,\Gamma_{\scriptscriptstyle WZ}[U,g])&\equiv&
\frac{{\cal W}^{-}(U,g)^{\ast}
{\cal W}^{+}(U,g)}{{\cal W}^{-}(U,{\bbone})^{\ast}{\cal W}^{+}(U,{\bbone})}\qp
\label{2}
\end{eqnarray}

If all the link variables are set to be unity,
it is expected that Eq.~(\ref{2}) gives the Wess-Zumino
effective action (we call as Wess-Zumino term in the following).
In this case the vacuum overlaps (\ref{3}) are expressed by
\begin{equation}
_{{\bbone}^g}\langle\pm |\pm\rangle_{\bbone}
=\mathop{\det}_{\scriptscriptstyle K,K'}
\lp\sum_n\psi^{\pm}_K(n,{\bbone})^{\dagger}g_n\psi^{\pm}_{K'}(n,{\bbone})\rp\qc
\label{4}
\end{equation}
and the Wess-Zumino term is defined by
\begin{eqnarray}
\exp(i\,\Gamma_{\scriptscriptstyle WZ}[g])&\equiv&
\frac{ {_{\bbone}}\langle -|-\rangle _{{\bbone}^g}
       {_{{\bbone}^g}}\langle +|+\rangle_{\bbone}  }
{ | {_{\bbone}}\langle -|-\rangle _{{\bbone}^g} |
  | {_{{\bbone}^g}}\langle +|+\rangle_{\bbone}  | }
\label{5}
\end{eqnarray}
For the continuum theory in two dimensions,
it is well-known that the chiral determinant can be
calculated exactly and the explicit expression can be obtained\cite{wz}.
Next we verify that Eq.~(\ref{5}) has the correct continuum limit
in two dimensions.

\subsection{Perturbative Calculation of Wess Zumino term }

First we examine (\ref{4}) perturbatively. The group elements $g_n$
can be expressed as $ e^{i\phi_n}$ by the variable $\phi_n$ which takes
the value on the Lie algebra. We take this
variables as the small quantum fluctuation from the unity
(not some background) so that we can expand as follows:
\begin{equation}
g_n=e^{i\,\phi_n}\cong {\bbone}+i\,\phi_n+\frac{1}{2!}(i\,\phi_n)^2+\cdots \qp
\end{equation}
Then, $\Gamma_{\scriptscriptstyle WZ}[g]$ can also be expanded
in $\phi_n$ and the continuum limit of the terms in the expansion series
can be evaluated.

Let $\Phi^{\pm}[g]$ denote the effective actions for Eq.~(\ref{4}).
Expanding by $\phi_n$, we obtain the following result,
\begin{eqnarray}
\Phi^{\pm}[g]&\equiv&\sum_{M=1}^{\infty}
\frac{(-1)^{M-1}}{M}\int_{k_1,\cdots,k_M}\tr
\lp{\widetilde {\delta g}}(k_1)\cdots
{\widetilde {\delta g}}(k_M)\rp\delta_{0,k_1+\cdots+k_M}\nonumber\\
&&\qquad\times C^{\pm}_{\scriptscriptstyle (M)}(k_1,\cdots,k_{M-1})\qc\\
C^{\pm}_{\scriptscriptstyle(M)}(k_1,\cdots,k_{M-1})
&=&\int_p O^{\pm}_{p,p-{K_1}}
O^{\pm}_{p-{K_1},p-{K_2}}\cdots
O^{\pm}_{p-{K_{M-1}},p}\nonumber\\
&=&\int_p\tr\lp S^{\pm}(p)S^{\pm}(p-{K_1})\cdots
S^{\pm}(p-{K_{M-1}})\rp\qc
\end{eqnarray}
where
\begin{eqnarray}
\int_k&\equiv&\int_{-\pi}^{\pi}\frac{d^2k}{(2\pi)^2}\qc\\
K_j&\equiv&\sum^j_{i=1}k_i,\quad K_0\equiv0\qc\\
(\delta g_n)_{ij}&\equiv&(g_n)_{ij}-\delta_{ij},\quad
{\widetilde {\delta g}}(k)\equiv\sum_n e^{-ik\cdot n}\delta g_n \qc\\
O^{\pm}_{pq}&\equiv&[\psi^{\pm}_p]^{\dagger}[\psi^{\pm}_q]\qc\\
S^{\pm}(p)_{\alpha\beta}
&\equiv&[\psi^{\pm}_p]_\alpha[\psi^{\pm}_p]^{\dagger}_\beta\qc\nonumber\\
&=&\frac{1}{2\lpm_p}\lp\lpm_p\delta_\ab+H^{\pm}(p)_\ab\rp\qc
\end{eqnarray}
$H^\pm(p)$ is the momentum representation of
Hamiltonians Eq.~(\ref{6}) in the free case
and $\psi^\pm_p $ and $\lambda^\pm_p$ are the eigenfunctions
and the eigenvalues of them.
The explicit expressions are given in the appendix A.
Then the Wess-Zumino term on the lattice is given by
\begin{eqnarray}
\gwzg={\rm Im}\lp\Phi^+[g]-\Phi^-[g]\rp\quad .
\end{eqnarray}
The oder $\phi$ contribution stems from $M=1$ term,
but $ C^+_{\scriptscriptstyle(1)}$ and $ C^-_{\scriptscriptstyle(1)}$
are equal to each other and cancel.
The oder $\phi^2$ contribution also vanishes since $\phi^2$ term in
$M=2$ terms are real. Therefore the leading term is the order $\phi^3$
and is given by the following,
\begin{eqnarray}
\hspace{-2cm}\gwzg^{(3)}&=&i\,\frac{1}{3}
\int_{k_1,k_2,k_3}\tr \lp {\widetilde \phi}(k_1)
{\widetilde \phi}(k_3){\widetilde \phi}(k_3) \rp
\delta_{0,k_1+k_2+k_3}\nonumber\\
&&\quad\times{\rm Im}\lp C^+_{\scriptscriptstyle(3)}(k_1,k_2)
                        -C^-_{\scriptscriptstyle(3)}(k_1,k_2)\rp \qp
\end{eqnarray}
where the explicit expressions of
$C^\pm_{\scriptscriptstyle (3)}$ are also given in the
Appendix A.
In order to take the continuum limit, we introduce physical momenta $p_i$ and
express the dimensionless momenta $k_i$ as $k_i=p_i a$ with the lattice
spacing $a$. Then we expands the coefficient
$C^\pm_{\scriptscriptstyle (3)}$ in $a$ and
obtain
\begin{eqnarray}
\hspace{-4cm}\gwzg^{(3)}&\cong&i\,\frac{1}{3}
\int_{p_1,p_2,p_3} \frac{d^2p_1}{(2\pi)^2}
\frac{d^2p_2}{(2\pi)^2}\frac{d^2p_3}{(2\pi)^2}
\tr\lp{\widetilde \phi}(p_1)
{\widetilde \phi}(p_3){\widetilde \phi}(p_3)\rp
\delta_{0,p_1+p_2+p_3}\nonumber\\
&&\qquad\qquad\qquad \times \left[ \lp\frac{1}{4\pi^2}
\epsilon_{\mu\nu}p_{1\mu}p_{2\nu}\rp\lp J^+-J^-\rp +O(a^2)
\right]\qc\label{7}\\
J^\pm&=&\int^\pi_{-\pi}d^2k\,\frac{1}{8{\lpm_k}^3}\left[
\lp C_\mu^{,\,\nu}(k)C_\nu^{,\,\mu}(k)-C_\mu^{,\,\nu}(k)^2\rp\lp
B(k)\mp m_0 \rp \right.
\nonumber\\
&&\quad\quad
\left. +2\lp C_\mu(k)C_\nu^{,\,\nu}(k)-C_\nu(k)C_\mu^{,\,\nu}(k)\rp
B(k)^{,\,\mu} \right]\qc
\label{eq:Jpm}
\end{eqnarray}
with $f^{,\,\mu}=\partial_\mu f$ .

For the Wilson fermion, $C$ and $B$ are given by
\begin{eqnarray}
C_\mu(k)=\sin k_\mu,\quad B(k)=\,r\sum_\mu \lp 1-\cos k_\mu \rp \qp
\end{eqnarray}
Then the integrands of $J^{\pm}$ are complex expressions of sine and
cosine  and include two parameters, $m_0$ and $r$. It dose not seem
to be able to accomplish the integrals analytically. By numerical
calculation we obtained $J^+=\pi,J^-=0$ for the region of
$m_0/r~, 0< m_0/r < 2$. Therefore, when $a$ goes to zero
Eq.~(\ref{7}) becomes as follows:
\begin{eqnarray}
\gwzg^{(3)}\mathop{\cong}_{a\rightarrow 0}-\frac{i}{12\pi}\int d^2x\,
\epsilon_{\mu\nu}\tr\lp\phi(x)\partial_\mu\phi(x)\partial_\nu\phi(x)\rp\qp
\end{eqnarray}
This is identical with the leading term of the continuum Wess-Zumino
term,
\begin{equation}
\gwzg^c=\frac{1}{12\pi}\int d^3y\,\epsilon^{ABC}
\tr\lp g^{-1}\partial_Ag\,g^{-1}\partial_Bg\,g^{-1}\partial_Cg\rp
\end{equation}
in the $\phi$-expansion.
\footnote{
For the continuum vacuum overlap formula,\cite{olnn}
the same calculation can be performed.
We obtain the similar expansion
as Eq.~(\ref{eq:Jpm})
in the $(k/m)$-expansion where m is Pauli-Villars mass.
\begin{eqnarray}
J^+=-J^-&=&\int d^2p\,\frac{m}{4\lambda_p^3}\qc\\
\lambda_p&\equiv&[p^2+m^2]^{\frac{1}{2}}\quad .\nonumber
\end{eqnarray}
It is easy to carry out this integral analytically and obtain the
results $J^+=-J^-=\pi/2$. It also reproduce the correct Wess-Zumino term.
The fact that $J^-=0$ in lattice regularization is consistent with the
calculation of the Chern-Simons current by Golterman et al.\cite{golt}
}
\subsection{Wess-Zumino-Witten model on the lattice}
\label{subsubsec:WZW}
Next we formulate a model with the Wess-Zumino term
defined in the previous subsections.
The partition function of the model is defined by
\begin{eqnarray}
Z=\int {\cal D}g\,e^{-S[g]}\,e^{in\gwzg}\qc
\end{eqnarray}
where $S[g]$ is the action of $SU(N)$ spin model:
\begin{eqnarray}
\label{eq:action_chiral_field}
S[g]&=&\Inc \sum_{n,\mu}\tr\lp\Hdel_\mu g(n)
\Hdel_\mu g(n)^\dagger\rp\nonumber\\
&=&-\Inc \sum_{n,\mu}\tr\lp g(n) g(n+\Hmu)^\dagger+g(n)^\dagger g(n+\Hmu)\rp
+{\rm const}\qc
\end{eqnarray}
and $n$ is an arbitrary integer. This model is considered as the
lattice counterpart of the two dimensional Wess-Zumino-Witten
model.\cite{witt}
In order to verify that the theory have the correct continuum limit,
we perform the calculation of the Callan-Symanzik $\beta$ function and
show it is identical to that of the continuum model.

To evaluate the $\beta$ function, one may use the background field
method, in which the field variables are separated into two parts
such as
\begin{eqnarray}
g_n=g_{0\,n}\,e^{\lambda i \pi_n}\qc
\label{8}
\end{eqnarray}
with $g_{0\,n}$ a smooth background and $\pi_n$ as small fluctuation
around $g_{0\,n}$. Then one calculates the correction to the
coefficient of the functional of $g_{0\,n}$,
\begin{eqnarray}
S_{cl}[g_0]=-\sum_{n,\mu}\tr\lp g_{0\,n}^{-1}{\widehat \partial}_\mu
g_{0\,n}\rp^2\qp
\label{9}
\end{eqnarray}
However, in our model, it is not easy to calculate the Wess-Zumino
functional defined by Eq.~(\ref{2}) keeping the nontrivial
background $g_{0\,n}$. So we first expand the total action
in $\phi_n$ by which $g_n$ is expressed as $e^{i\phi_n}$.
Then we shift $\phi_n$ around a background $\phi_{0\,n}$
not linearly but nonlinearly by applying the Hausdorff's formula to
Eq.~(\ref{8}) as follows:
\begin{eqnarray}
\phi_n&=&\phi_{0\,n}+\lambda\pi_n
+\frac{i}{2}\lambda[\phi_{0\,n},\pi_n]\nonumber\\
&&-\frac{1}{12}\lp\lambda[\phi_{0\,n},[\phi_{0\,n},\pi_n]\,]
+\lambda^2[\pi(n),[\pi(n),\phi_{0\,n}]\,]\rp+\cdots\qp
\end{eqnarray}
In this case, the lowest order of the functional (\ref{9}) is
\begin{equation}
S_{cl}[\phi_0]=\sum_{n,\mu}\tr\lp{\widehat \partial}_\mu\phi_{0\,n}
{\widehat \partial}_\mu \phi_{0\,n}\rp\qc
\end{equation}
and it is sufficient for us to estimate the corrections to
its coefficient.
The one-loop order corresponds to $O(\phi_0^2,\lambda^2)$.
To obtain the propagator and vertices necessary
to calculate the correction to this order,
we first expand the actions $S[g]$
and $\Gamma_{\scriptscriptstyle WZ}[g]$ in $\phi_n$ as follow,
\begin{eqnarray}
S[g]&=&\Inc\sum_{n,\mu}\tr ( \lp \Hdel_\mu\phi_n\rp^2
-\frac{1}{12}\Hdel_\mu\phi_n [ \phi_n,[
\phi_n,\Hdel_\mu\phi_n]] \nonumber \\
&&\hspace{3cm}-\frac{1}{12}\lp \Hdel_\mu\phi_n\rp^4 +\cdots )\qc\\
\gwzg&=&\int_{k_1,k_2,k_3}{\widetilde \phi^{a_1}}(k_1)
{\widetilde \phi^{a_2}}(k_2){\widetilde \phi^{a_3}}(k_3)
\delta_{0,k_1+k_2+k_3}\nonumber\\
&&\times V_{wz}^{(3)a_1,a_2,a_3}(k_1,k_2)+ \cdots\qc\\
V_{wz}^{(3)a_1,a_2,a_3}(k_1,k_2)&=&-\frac{1}{12}f^{a_1 a_2 a_3}
{\rm Im}(C_{\scriptscriptstyle (3)}^+ (k_1,k_2)
-C_{\scriptscriptstyle (3)}^- (k_1,k_2)) .
\end{eqnarray}
Next we shift $\phi_n$ around $\phi_{0\,n}$ as mentioned above
and obtain the followings.
\begin{eqnarray}
S_0[\phi]&=&\Inc S_{cl}[\phi] \qc\\
S_{cl}[\phi]&=&\frac{1}{2}\int_k {\widetilde \phi}^a(-k)
{\widetilde \phi}^a(k)\Delta(k)^{-1} \qc
\end{eqnarray}
where
\begin{equation}
\Delta(k)^{-1}= \, \sum_\mu \lp 1-\cos k_\mu \rp ,
\end{equation}
and
\begin{eqnarray}
S_1&=&\int_{k_1,k_2,k_3}
{\widetilde \phi_0}^{a_1}(k_1) {\widetilde \pi}^{a_2}(k_2)
{\widetilde \pi}^{a_3}(k_3)\delta_{0,k_1+k_2+k_3}\nonumber\\
&&\times V_1^{a_1,a_2,a_3}(k_3)\qc\\
S_2&=&\int_{k_1,k_2,k_3,k_4}
{\widetilde \phi_0}^{a_1}(k_1) {\widetilde \phi_0}^{a_3}(k_3)
{\widetilde \pi}^{a_2}(k_2)
{\widetilde \pi}^{a_4}(k_4)\delta_{0,k_1+k_2+k_3+k_4}\nonumber\\
&&\times V_2^{a_1 a_2 a_3 a_4}(k_1,k_2)\qc\\
S_3&=&\int_{k_1,k_2,k_3,k_4}
{\widetilde \phi_0}^{a_1}(k_1) {\widetilde \phi_0}^{a_2}(k_2)
{\widetilde \pi}^{a_3}(k_3)
{\widetilde \pi}^{a_4}(k_4)\delta_{0,k_1+k_2+k_3+k_4}\nonumber\\
&&\times (V_3^{a_1 a_2 a_3 a_4}(k_3,k_4)
+V_3^{a_3 a_4 a_1 a_2}(k_1,k_2)+4 V_3^{a_3 a_2 a_1 a_4}(k_1,k_4)) \qc \\
S_4&=&\int_{k_1,k_2,k_3,k_4}
{\widetilde \phi_0}^{a_1}(k_1) {\widetilde \phi_0}^{a_2}(k_2)
{\widetilde \pi}^{a_3}(k_3)
{\widetilde \pi}^{a_4}(k_4)\delta_{0,k_1+k_2+k_3+k_4}\nonumber\\
&&\times (4 V_4^{a_1 a_2 a_3 a_4}(k_1,k_2,k_3,k_4)
+2 V_4^{a_2 a_1 a_3 a_4}(k_1,k_2,k_3,k_4)),
\end{eqnarray}
\begin{eqnarray}
\Gamma_{\scriptscriptstyle wz,1}
&=&3\lambda^2\int_{k_1,k_2,k_3}
{\widetilde \phi_0}^{a_1}(k_1) {\widetilde \pi}^{a_2}(k_2)
{\widetilde \pi}^{a_3}(k_3) \delta_{0,k_1+k_2+k_3}\nonumber\\
&&\times V^{(3) a_1 a_2 a_3}_{\scriptscriptstyle wz}(k_1,k_2)\qc\\
\Gamma_{\scriptscriptstyle wz,2}
&=&-3\lambda^2\int_{k_1,k_2,k_3}
{\widetilde \phi_0}^{a_1}(k_1)
{\widetilde \pi}^{a_2}(k_2) {\widetilde \phi_0}^{a_3}(k_3)
{\widetilde \pi}^{a_3}(k_3) \delta_{0,k_1+k_2+k_3+k_4}\nonumber\\
&&\times V^{(3) a_1 a_2 e}_{\scriptscriptstyle wz}(k_1,k_2)f^{a_3,a_4 e}\qp
\end{eqnarray}
The vertex functions $V$'s are listed in appendix B.
Then the partition function is calculated as follows:
\begin{eqnarray}
Z = e^{-S_0[\phi_0]}&&\hspace{-0.4cm}
\int {\cal D}\pi e^{-S_0[\pi]}\nonumber\\
&& \times
e^{-S_1 -S_2 -S_3 -S_4
   + i\,n\Gamma_{\scriptscriptstyle wz,1}
   + i\,n\Gamma_{\scriptscriptstyle wz,2} }\\
= - S_0[\phi_0]&+&\frac{1}{2!}
\langle(S_1)^2\rangle_0
-\langle S_2\rangle _0
-\langle S_3\rangle _0-\langle S_4\rangle_0\nonumber\\
-\frac{(\lambda^2 n)^2}{2!}
&&
\hspace{-0.4cm}
\langle (\Gamma_{\scriptscriptstyle wz}^{(1,2)})^2\rangle_0
-i\,n\lambda^2\langle \Gamma_{\scriptscriptstyle wz}^{(2,2)}\rangle_0\qc
\end{eqnarray}
where
\begin{eqnarray}
\langle O \rangle_0 \equiv \int {\cal D}\pi O(\pi)e^{-S_{2,0}^{(0,2)}[\pi]}\qp
\end{eqnarray}
The contribution of the original spin model is calculated as\cite{kog}
\begin{eqnarray}
\frac{C_A}{4}\Delta(0)S_{cl}[\phi_0]\mathop{\simeq}_{a\rightarrow 0}
-\lp \frac{C_A}{4}\frac{1}{4\pi}\ln a^2 m^2 \rp S_{cl}[\phi_0] \qc
\end{eqnarray}
where $C_A$ is the seconde Casimir coefficient in the adjoint
representation and $m$ is the infrared cut-off mass. On the other hand,
the contribution from the Wess-Zumino term is calculated
as follows. It is easy to see the second term vanishes.
The first term is given by
\begin{equation}
-\frac{(\lambda^2 n)^2}{2!}\langle
(\Gamma_{\scriptscriptstyle wz}^{(1,2)})^2\rangle_0
=\frac{1}{2}\int_k {\widetilde \phi}^a_0(-k){\widetilde \phi}^a_0(k)
\lp I_1(k)+I_2(k)\rp\qc
\end{equation}
\begin{eqnarray}
&&I_1(k)=-\lp\frac{\lambda^2 n}{4}\rp^2 C_A
\int_p\Delta(p)\Delta(k-p)C(k,-p)^2 \qc\\
&&I_2(k)=\lp\frac{\lambda^2 n}{4}\rp^2 C_A
\int_p\Delta(p)\Delta(k-p)C(k,-p)C(-k,k-p)\qc \\
&&C(k,p)={\rm Im} \lp C_{\scriptscriptstyle (3)}^+(k,p)
-C_{\scriptscriptstyle (3)}^+(k,p) \rp\qp
\end{eqnarray}

Toward the continuum limit we also scale the dimensionless external momenta $k$
by $a$ and consider the momenta $ka$ near zero. Following
Karsten and Smit\cite{kar}, we can estimate
the loop integral. The divergent contribution stems
from the momenta near zero, so that we can use the continuum limit
of the vertex $C(k,p)$ obtained in the previous section. It is
calculated as
\begin{eqnarray}
I_1(k)\sim I_2(k)\sim \frac{1}{(4\pi)^2}\lp\frac{\lambda^2 n}{4}\rp^2
C_A\frac{k^2}{4\pi}\ln a^2 k^2\qp
\end{eqnarray}
Therefore the contribution from Wess-Zumino term is
\begin{eqnarray}
-\frac{(\lambda^2 n)^2}{2!}\langle
(\Gamma_{\scriptscriptstyle wz}^{(1,2)})^2\rangle_0
\mathop{\simeq}_{a\rightarrow 0}\lp \frac{\lambda^2 n}{8\pi}\rp^2 \frac{C_A}{4}
\lp\frac{1}{4\pi} \ln a^2 \rp S_{cl}[\phi_0]+{\rm const} \qp
\end{eqnarray}
Adding two parts, we obtain
\begin{equation}
Z\mathop{\simeq}_{a\rightarrow 0}
\exp\lp -\frac{1}{\lambda_R^2}S_{cl}[\phi_0]+\cdots \rp\qc\\
\end{equation}
where $\lambda_R^2$ denotes the renormalized coupling constant given by
\begin{eqnarray}
\frac{1}{\lambda_R^2}=\frac{1}{\lambda^2}
+\lp 1-\lp\frac{n\lambda^2}{8\pi}\rp ^2\rp
\frac{C_A}{4}\frac{1}{4\pi}\ln a^2 \qp
\end{eqnarray}
{}From this, $\beta$ function is obtained as
\begin{equation}
\beta(\lambda^2)=-a\frac{\partial\lambda^2}{\partial a}=
-\lp 1-\lp\frac{n\lambda^2}{8\pi}\rp^2\rp\frac{C_A}{4}
\lp\frac{\lambda^4}{2\pi}\rp .
\end{equation}
For $n >0$, the $\beta$ function has an IR-fixed point
at $\lambda^2=\frac{8\pi}{n}$
as expected (the factor 2 differ from Witten's results\cite{witt}
because of the difference of the convention).
For $\lambda^2 < \frac{8\pi}{n}$, this model is asymptotically free,
so that it has a renormalization group invariant scale parameter,
$\Lambda$ given by
\begin{eqnarray}
\Lambda=\frac{1}{a}
\left|\frac{\lambda_c^2+\lambda^2}{\lambda_c^2-\lambda^2}
\right|^{\frac{1}{2\beta_0\lambda_c^2}}
\exp\lp -\frac{1}{\beta_0\lambda^2}\rp\qc\\
\beta_0=\frac{C_A}{8\pi}\qc \quad \lambda_c^2=\frac{8\pi}{n}\qp
\end{eqnarray}

\section{Numerical Calculation of the Wess-Zumino term}

In this section, we show our results of
the numerical calculation of the Wess-Zumino term.
We consider the SU(2) chiral field and adopt
the lattice action with the nearest neighbor coupling.
We generate SU(2) chiral fields by the cluster algorithm.\cite{cluster}
Lattice size is set to 16 x 16.
For most of the values of $\beta \equiv \frac{2}{\lambda^2}$ examined,
500 configurations are generated. For $\beta=1.1, 1.5$,
we generate 4000 more configurations.
With these configurations, the vacuum overlaps are calculated
numerically through the LU decomposition.
We measure the distribution of the Wess-Zumino term
by dividing the range of the value $[-\pi,\pi]$ into bins with the
width 0.1.
We also calculate the standard deviation $\sigma_{WZ}$ of the distribution
of the Wess-Zumino term for each $\beta$.

\begin{figure}
\epsfysize=6cm
\centerline{\epsffile{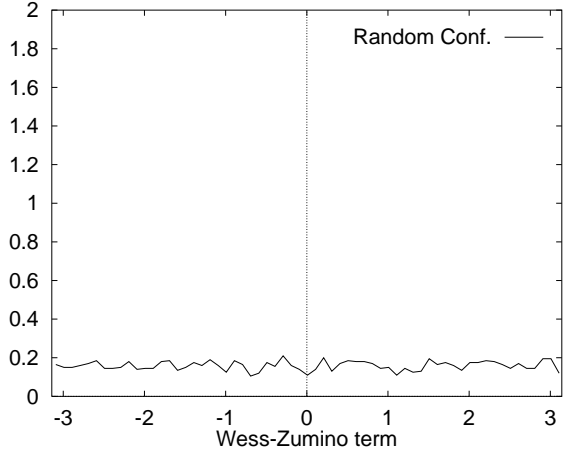}}
\caption{
The distribution of the two-dimensional Wess-Zumino term
for randomly generated SU(2) chiral fields.
The range of the value $[-\pi,\pi]$ is divided into bins
by the width 0.1.
The normalized number of configurations for each bin is shown.
L=16.
}
\label{fig:random_conf_dist}
\end{figure}

\begin{figure}
\epsfysize=6.5cm
\centerline{
\epsffile{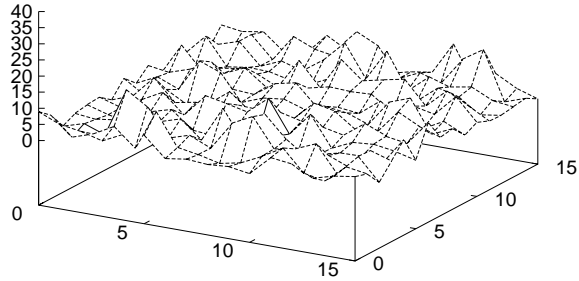}
}
\caption{
A typical configuration of random SU(2) chiral field
in momentum space: the absolute value of the trace part is shown.
L=16.
}
\label{fig:random_conf_u0}
\end{figure}

First of all,
we show the behavior of the Wess-Zumino term for randomly
generated configurations of SU(2) chiral field
in Figure \ref{fig:random_conf_dist}.
($\beta$ is set to $0.025$.) The Wess-Zumino term distributes
almost uniformly in the range $[-\pi,\pi]$.
In Figure \ref{fig:random_conf_u0}, we show
a typical configuration of the random SU(2) chiral field in momentum space.
We can find that almost all modes of momenta contribute equally.
The species doublers are then suspected to contribute to the complex phase.


\begin{figure}
\epsfysize=6cm
\centerline{
\epsffile{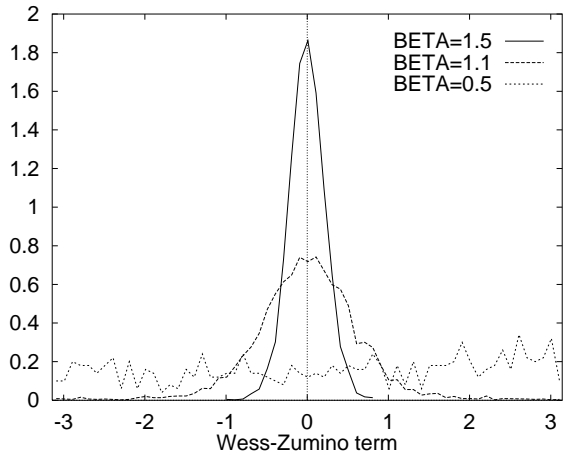}
}
\caption{
The distribution of the two-dimensional Wess-Zumino term
for $\beta=0.5$, $1.1$ and $1.5$.
}
\label{fig:gauss}
\end{figure}

\begin{figure}
\epsfysize=6cm
\centerline{
\epsffile{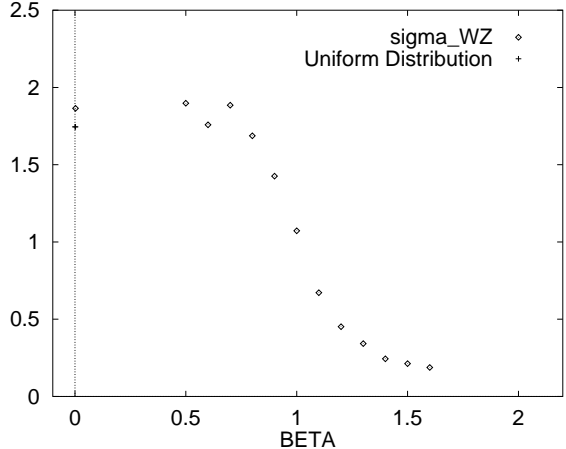}
}
\caption{The standard deviation $\sigma_{WZ}$ of the distribution
of the Wess-Zumino term for various $\beta$.}
\label{fig:sigma}
\end{figure}

Next we show the distributions of the Wess-Zumino term for
$\beta = 0.5$, $1.1$ and $1.5$ in Figure \ref{fig:gauss}.
We can clearly observe that
for $\beta=1.1$, the evaluated values of the Wess-Zumino term
start to concentrate around zero and the distribution becomes
like Gaussian.
When $\beta$ increases, the distribution becomes sharp Gaussian.
$\beta$ dependence of the standard
deviation $\sigma_{WZ}$ of the Wess-Zumino term distribution
is plotted in Figure \ref{fig:sigma}.
The onset of the Gaussian distribution is around $\beta=0.9$.
Crossover is sharp from the strong coupling region
where the Wess-Zumino term fluctuates hard.

\begin{figure}
\epsfysize=7cm
\centerline{
\epsffile{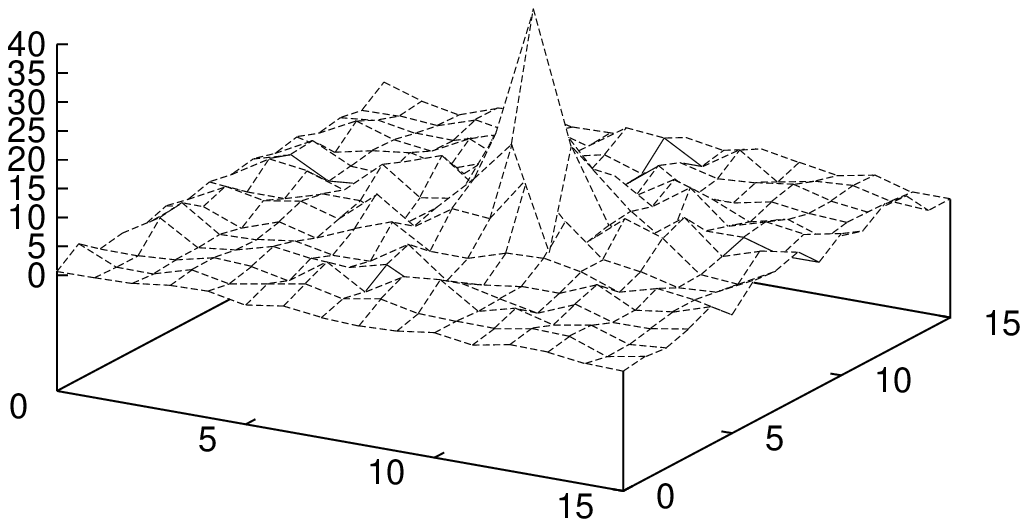}
}
\caption{
A typical configuration of SU(2) chiral field at $\beta=1.5$
in momentum space: the absolute value of the trace part is shown.
L=16.
}
\label{fig:smooth_conf_u0}
\end{figure}

In Figure \ref{fig:smooth_conf_u0}, we show
a typical configuration of the SU(2) chiral field at $\beta=1.5$
in momentum space.
We can find that the modes with small momenta dominates.
In this region of $\beta$, the species doublers are almost suppressed.

\begin{figure}
\epsfysize=6cm
\centerline{
\epsffile{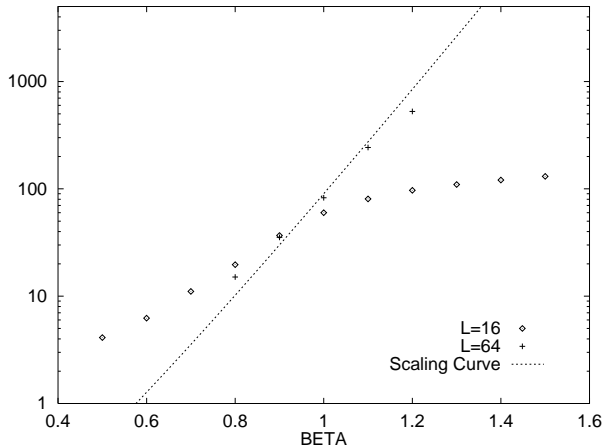}
}
\caption{
Scaling of susceptibility of SU(2) chiral model.
L=16 and L=64.
}
\label{fig:scaling_chi}
\end{figure}

In Figure \ref{fig:scaling_chi}, we show the susceptibility $\chi$
of the SU(2) chiral model for $L=16$ and $L=64$.
The scaling of the susceptibility starts around $\beta=0.9$.
This coincides with the onset of the Gaussian distribution
of the Wess-Zumino term.

In summary,
the two-dimensional Wess-Zumino term defined by the vacuum
overlap formula shows a sharp Gaussian distribution
for the configurations of SU(2) chiral field in the scaling region.
In this region, the configurations are smooth and the contribution
of the species doublers is suppressed.
In the strong coupling region,
the Wess-Zumino term fluctuates hard and the species
doublers' contribution is suspected to affect it.
However the crossover is rather sharp from the strong coupling
region to the weak coupling region.

A few comments are in order for our numerical calculation.
In Figure \ref{fig:scaling_chi}, we find that the finite size
effect on $\chi$ is substantial for $L=16$.
We also suspect the finite size effect on the Wess-Zumino term.
The onset of the Gaussian distribution of the Wess-Zumino term in the
scaling region is also expected for larger lattice.
However there can be much finite size effect in the values
of $\sigma_{WZ}$. We need further investigation on large lattice.

\section{Numerical Evaluation of Observable in Wess-Zumino-Witten model}

In this section, we discuss the
possibility of the numerical estimate of observables
in the Wess-Zumino-Witten model by incorporating the Wess-Zumino action.
The lattice Wess-Zumino-Witten model we are considering
is defined with the complex action,
\begin{equation}
  \label{action_WZW}
  S_{WZW}[g]=S[g] - i n \Gamma_{WZ}[g] ,
\end{equation}
where $S[g]$ and $\Gamma_{WZ}[g]$ are given
by Eq.~(\ref{eq:action_chiral_field}) and by Eq.~(\ref{5}), respectively.
$n$ is the integer coupling constant of the Wess-Zumino term.
In order to evaluate the observables in this model,
we need to perform the following integral,
\begin{eqnarray}
  \label{eq:integral_WZW}
\langle O \rangle &=&
\int {\cal D}g\,e^{-S[g]+ i n \gwzg} O[g] \, \big/ Z , \\
Z&=& \int {\cal D}g\,e^{-S[g]+ i n \gwzg} .
  \label{eq:integral_WZWb}
\end{eqnarray}

In general, without the positivity of the measure,
the Monte Carlo method cannot be applied.
One possible way out is to incorporate the
imaginary Wess-Zumino term into the observable and to perform
the Monte Carlo integration only with the real part of the action.
That is, we consider
\begin{eqnarray}
  \label{eq:integral_WZW_MC}
\langle O \rangle &=&
\langle\langle e^{i n \gwzg} O[g] \rangle\rangle_{MC} \, \big/ Z , \\
Z&=& \langle\langle e^{i n \gwzg} \rangle\rangle_{MC},
\end{eqnarray}
where $\langle\langle \, \rangle\rangle_{MC}$ denotes the Monte Carlo
integration only with the real part of the action,
\begin{equation}
  \label{eq:integral_MC}
\langle\langle X \rangle\rangle_{MC} \equiv
\int {\cal D}g\,e^{-S[g]} X[g] .
\end{equation}

Another possible method is to introduce
the spectral functions\cite{gock} defined as follows:
\begin{eqnarray}
  \label{eq:spectral_func}
\rho (\theta)&\equiv&
\int {\cal D}g e^{-S[g]} \delta (\theta-\gwzg) , \\
O(\theta)&\equiv&\int {\cal D}g e^{-S[g]} \delta(\theta-\gwzg) O(g), \\
\end{eqnarray}
where the delta function is the periodic one.
With these spectral functions, we can write the integrals
Eq.~(\ref{eq:integral_WZW}) and Eq.~(\ref{eq:integral_WZWb}) as
\begin{eqnarray}
  \label{eq:integral_WZW_spectral}
\langle O \rangle &=&
\int_{-\pi}^\pi d\theta  \, e^{i\,n\,\theta } O(\theta ) \, \big/ Z ,\\
Z &=& \int_{-\pi}^\pi d\theta  \, e^{i\,n\,\theta } \rho(\theta).
\end{eqnarray}
The integration of $\theta$ is then performed as the Riemann integral.

Actually, the distribution of the Wess-Zumino term obtained in the
previous section can be regarded as the spectral function,
\footnote{
It may be better to use the set method\cite{set-method}
for the precise evaluation of the spectral function.
In this case, however, since the value of the Wess-Zumino term is
in the compact region $[-\pi,\pi]$, we can do without it if we have
enough statistics.
}
\begin{equation}
  \label{eq:spectral_func_WZ_MC}
\rho(\theta)=
\langle\langle
\delta(\theta -\Gamma_{WZ}) \rangle\rangle_{MC} ,
\end{equation}
in the discrete approximation with the normalization,
\begin{equation}
  \label{normalization_spectral_func_WZ_MC}
  \sum_i \Delta\theta \, \rho(\theta_i) = 1 .
\end{equation}
We found that in the scaling region, it is given
by a sharp Gaussian distribution with the width $\sigma_{WZ}$
depending on $\beta$,
\begin{equation}
  \label{gaussian_WZ}
\rho(\theta_i) \sim
\frac{1}{\sqrt{2\pi} \sigma_{WZ}} \sum_{m=-\infty}^{\infty}
e^{
- \frac{1}{2 \sigma_{WZ}^2} ( \theta_i+ 2\pi m )^2
} .
\end{equation}

\begin{figure}
\epsfysize=6cm
\centerline{\epsffile{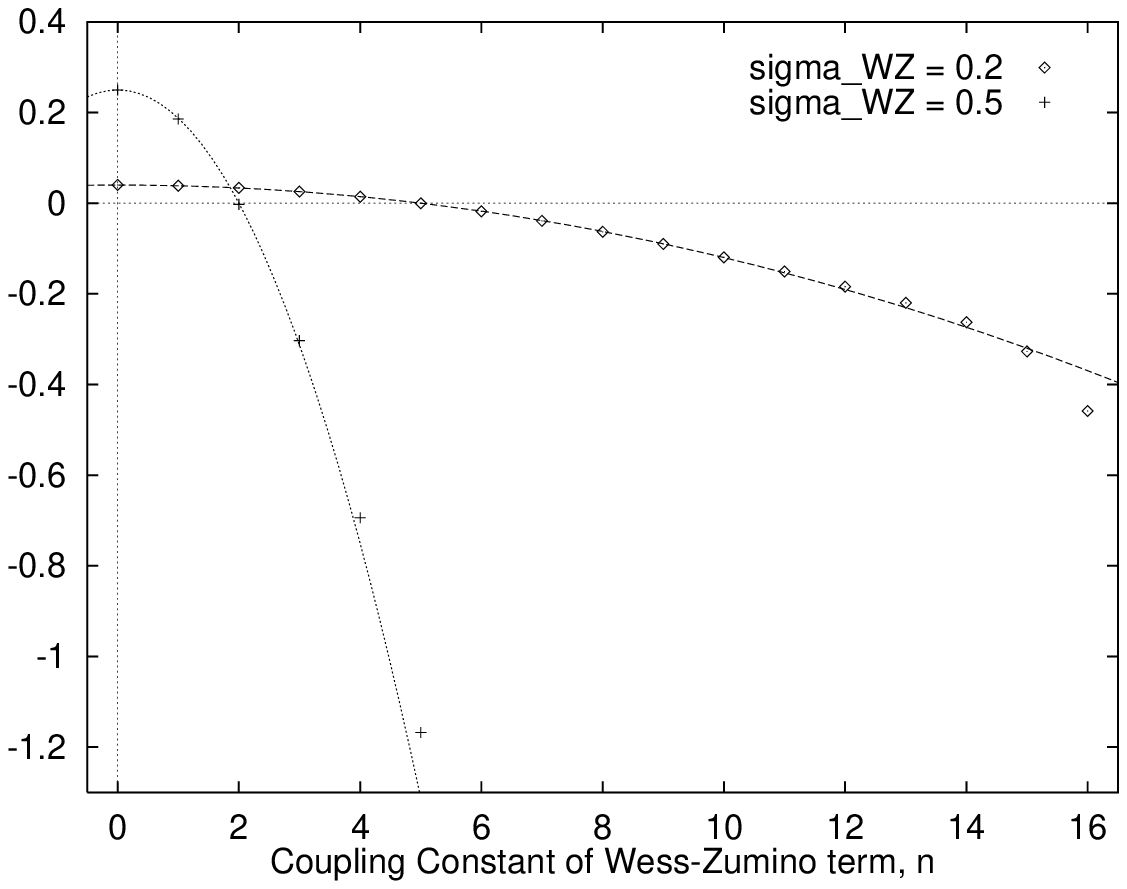}}
\caption{Monte Carlo integration of $\theta^2 \exp\left(i n \theta
  \right)$ with the Gaussian weight. The width of the Gaussian weight
$\sigma_{WZ}$ is chosen as $0.2$ and $0.5$.
The approximated analytical result for these $\sigma_{WZ}$,
$\sigma_{WZ}^2(1-\sigma_{WZ}^2 n^2)$, are
also shown.
}
\label{fig:theta_MC_x2}
\end{figure}

For such a Gaussian distribution of the imaginary part of
the action, it is known that
the first method mentioned above may work
only for small $n$.\cite{gock}
To see this, we perform the following test.
We assume the Gaussian distribution for the Wess-Zumino term,
Eq.~(\ref{gaussian_WZ}).
We take the square of the Wess-Zumino term as an observable.
Since the spectral function is obtained by the Monte Carlo method,
we simulate the situation by performing the Monte Carlo integration on
$\theta$. Thus we consider the following integrals for various $n$ at
fixed $\sigma_{WZ}$,
\begin{eqnarray}
  \label{eq:integral_WZW_MC_theta}
\langle (\Gamma_{WZ})^2 \rangle &=&
\langle\langle e^{i n \theta} \theta^2
\rangle\rangle_{\theta} \, \big/ Z , \\
Z&=& \langle\langle e^{i n \theta} \rangle\rangle_{\theta},
\end{eqnarray}
where $\langle\langle \, \rangle\rangle_{\theta}$ denotes the Monte Carlo
integration of $\theta$ with the weight $\rho(\theta)$,
\begin{equation}
  \label{eq:integral_MC_theta}
\langle\langle X \rangle\rangle_{\theta} \equiv
\int_{-\pi}^\pi d\theta  \, \rho(\theta) X[\theta] .
\end{equation}
In Figure \ref{fig:theta_MC_x2}, we show the result.
We found that for $\sigma_{WZ}=0.5$, up to $n=3$ the numerical result
reproduces the exact one within the acceptable error of
three significant digits.
For $\sigma_{WZ}=0.2$, $n=12$.

\begin{figure}
\epsfysize=6cm
\centerline{\epsffile{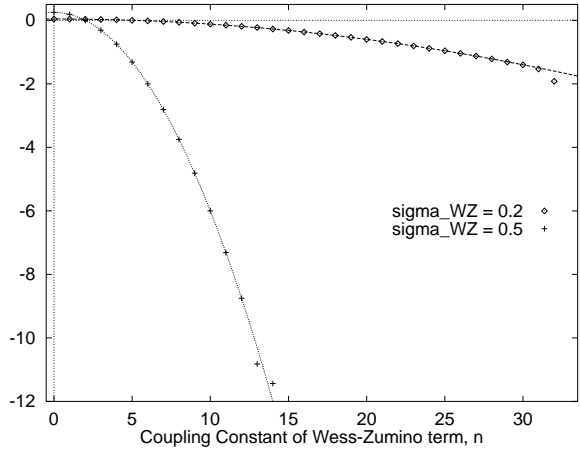}}
\caption{Riemann integration of $\theta^2 \exp\left(i n \theta
  \right)$ with the Gaussian weight. The width of the Gaussian weight
$\sigma_{WZ}$ is chosen as $0.2$ and $0.5$.
The approximated analytical result for these $\sigma_{WZ}$,
$\sigma_{WZ}^2(1-\sigma_{WZ}^2 n^2)$, are also shown.
}
\label{fig:theta_R_x2}
\end{figure}

On the other hand, it is also possible to perform the Riemann integral
in Eq.~(\ref{eq:integral_WZW_spectral}) numerically for the same
observable.
We also found the upper bounds for $n$.
In order that
the numerical result reproduces the exact one within the acceptable error,
we should have $n \le 12$ for $\sigma_{WZ}=0.5$
and $n \le 30$ for $\sigma_{WZ}=0.2$.
We have tried several schemes of higher order integration
and obtained the similar result.

\begin{figure}
\epsfysize=6cm
\centerline{\epsffile{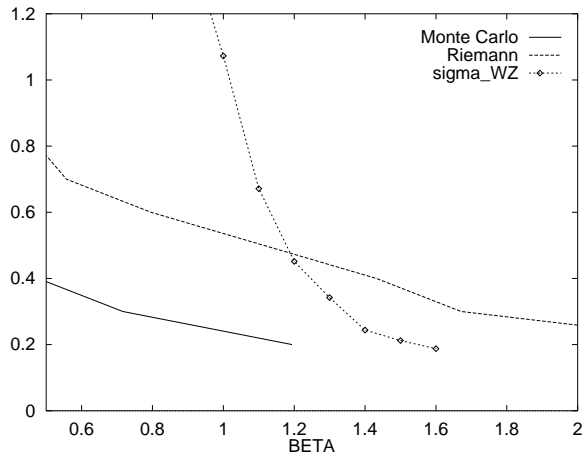}}
\caption{
Upper bound on $\sigma_{WZ}$ for given coupling constant $n$
of the Wess-Zumino term. $n$ is translated to the value of
the IR fixed point by the relation $\beta_c=\frac{n}{4\pi}$.
The values of $\sigma_{WZ}$ obtained in the previous section are
also plotted.
}
\label{fig:upperbound}
\end{figure}

Thus we obtained the upper bound for the accessible value of
the coupling constant $n$ of the Wess-Zumino term for given
$\sigma_{WZ}$.
In other word,
for a given $n$, there exists the upper bound of $\sigma_{WZ}$,
with which the measurement of the observable can be
performed by the direct Monte Carlo integration
or by the Riemann integration of the spectral functions.
In Figure \ref{fig:upperbound}, we show the upper bounds of $\sigma_{WZ}$
versus the coupling constant $n$ of the Wess-Zumino term which is
translated to the IR fixed point $\beta_c=\frac{n}{4\pi}$.
{}From this figure,
it seems possible to simulate the model across the IR fixed point
for $\beta_c \ge 1.3$, (that is $n \ge 17$),
if the spectral density method is used.
On the contrary, by the direct Monte Carlo calculation,
only the weak coupling region can be simulated.

Our discussion given here is based on the numerical result
shown in the previous section. Since the systematic and statistical
errors are suspected in the result, the numbers of the upper bounds
given in this discussion should not be taken literary.
We believe, however, that this discussion gives a correct qualitative
estimate of the possibility.

\section{Discussion and Conclusion}

The vacuum overlap formula for the two-dimensional Wess-Zumino term
was examined both perturbatively and nonperturbatively.
We showed by the lattice perturbation theory
that the formula correctly reproduces the Wess-Zumino term
in the continuum limit and yields the IR fixed point in the
beta function of the SU(2) chiral model.
We calculated the vacuum overlaps numerically and found that
the complex phase shows a sharp Gaussian distribution
for the SU(2) chiral fields in the scaling region.
In this region, the smooth configurations were
obtained even in the symmetric phase due to asymptotic freedom.
We also found that the crossover is rather sharp from the strong coupling
region where the Wess-Zumino term fluctuates hard and the species
doublers' contribution is suspected to affect it.

It was also argued that,
if we use the spectral density method,
it seems possible to examine the region of the
IR fixed point of the Wess-Zumino-Witten model numerically.

Also from this study, we found that
the asymptotically free coupling for gauge degree of freedom is able
to reduce the gauge fluctuation. In two dimensions, the nearest neighbor
coupling of the SU(2) chiral field plays such a role.
In four dimensions, it is known in the continuum theory that
the four-derivative coupling which is induced from the gauge
fixing term is asymptotically free.\cite{pure-gauge-model}
It may be interesting to examine the gauge degree of
freedom of the four-dimensional nonabelian gauge field with
the gauge fixing term from this point of view.\cite{shamir_gaugefix_local}

\section*{Acknowledgements}
The authors would like to thank H.~Hata for enlightening discussions
about the perturbative treatment of the non-linear sigma model.
They are also grateful to W.~Bock for his helpful advice about
the Monte Carlo simulation with the Cluster algorithm.

This work is supported by National Laboratory for High Energy Physics,
as KEK Supercomputer Project (Project No. 10).

\appendix
\section{Eigenvalues and Eigenfunctions of $H^\pm(p)$}
In this appendix we give the detail expression of the functions used
in subsection \ref{subsubsec:WZterm}
The momentum representation of Hamiltonians and the Wave functions are

\begin{eqnarray}
H^\pm(q)\psi_p^\pm(q)&=&\lpm_p \psi_p^\pm(q)\qc\\
H_{\ab}^\pm(p)&=&\lp
\begin{array}{cc}
B(p)\mp m_0 & C(p)\\
C(p)^\dagger   & -B(p)\pm m_0
\end{array}\rp\qc\\
\psi_p^\pm(q)_\alpha&=&\delta_{p,q}{\psi^\pm}_{p,\alpha}\\
\psi^\pm_{p,\alpha}&=&\frac{1}{N^\pm_p}\lp
\begin{array}{c}
\lpm_p+ B(p)\mp m_0 \\
C^\dagger(p)
\end{array}\rp\qc\\
\lpm_p&=&\left[ C_\mu(p)^2+\lp B(p)\mp m_0 \rp^2 \right]^\frac{1}{2}\qc\\
N^\pm_p&=&\left[ 2 \lpm_p \lp \lpm_p +B(p)\mp m_0 \rp \right]^\frac{1}{2}\qp
\end{eqnarray}
For Wilson fermion ,
\begin{eqnarray}
C_\mu(p)=\sin p_\mu , B(p)=\,r\sum_\mu \lp 1-\cos k_\mu \rp \qp
\end{eqnarray}
But one may take any regularization function as $C(p)$ and $B(p)$.
Imaginary part of $C_{\scriptscriptstyle (3)}^\pm$ are
\begin{eqnarray}
{\rm Im}C_{\scriptscriptstyle (3)}^\pm (k_1,k_2)
&=&\int_p \frac{B_{\scriptscriptstyle (3)}^\pm(p;k_1,k_2)}
{4 \lpm_p\lpm_{p-k_1}\lpm_{p-k_1-k_2}}\qc\\
B_{\scriptscriptstyle (3)}^\pm(p;k_1,k_2)&=&-\epsilon_{\mu\nu}[
C_\mu(p)C_\nu(p-k_1)\lp B(p-k_1-k_2)\mp m_0 \rp\nonumber\\
&&+\quad C_\mu(p-k_1)C_\nu(p-k_1-k_2)\lp B(p)\mp m_0 \rp\nonumber\\
&&+\quad C_\mu(p-k_1-k_2)C_\nu(p)\lp B(p-k_1)\mp m_0 \rp ]\qp
\end{eqnarray}
Expanding on $k_1,k_2$, one obtain the expression of the integral
$J^\pm$.

\section{Vertices for One-Loop Calculation}
Here we give the vertex functions used in
subsectection \ref{subsubsec:WZW}
\begin{eqnarray}
&&V_1^{a_1,a_2,a_3}(k_3)=-\frac{1}{2}f^{a_1 a_2 a_3}
\Delta(k_3)^{-1}\qc\\
&&V_2^{a_1 a_2 a_3 a_4}(k_1,k_2)=f^{a_1 a_2 e}f^{a_3 a_4 e} \nonumber\\
&&\qquad\qquad\qquad
\times\left(\frac{1}{8}\Delta(k_1+k_2)^{-1}
-\frac{1}{12}\lp\Delta(k_1)^{-1}+\Delta(k_2)^{-1}\rp \right)\qc\\
&&V_3^{a_1 a_2 a_3 a_4}(k,k')=-\frac{1}{12}f^{a_1 a_3 e}f^{a_2 a_4 e}
\, \sum_\mu (e^{i\,k_\mu}-1)(e^{i\,{k'}_\mu}-1)\qc\\
&&V_4^{a_1 a_2 a_3 a_4}(k_1,k_2,k_3,k_4)=-\frac{1}{12}
\tr \lp T^{a_1}T^{a_2}T^{a_3}T^{a_4}\rp\nonumber\\
&&\qquad\qquad\qquad
\times\, \sum_\mu (e^{i\,k_{1,\mu}}-1)(e^{i\,k_{2,\mu}}-1)
(e^{i\,k_{3,\mu}}-1)(e^{i\,k_{4,\mu}}-1)\qc\\
&&V^{(3) a_1 a_2 a_3}_{\scriptscriptstyle wz}(k_1,k_2)
=-\frac{1}{12}f^{a_1 a_2 a_3}
{\rm Im}\lp C_{\scriptscriptstyle (3)}^+(k_1,k_2)
           -C_{\scriptscriptstyle (3)}^+(k_1,k_2) \rp\qp
\end{eqnarray}

\section{Convention for SU(N) Lie Algebra}
\begin{eqnarray}
\tr \lp T^a T^b \rp &=& \frac{1}{2}\delta_{a\,b}\qc\\
 \left[ T^a, T^b \right ]&=&i\,f^{a b c} T^c \qc\\
\tr \lp T_{Ad}^a T_{Ad}^b \rp &=& f^{a c d}f^{b c d}=
C_A\delta_{a\,b}\qc\\
\phi(x)&=&\phi^a(x) T^a\qp
\end{eqnarray}


\begin{thebibliography}{99}
\bibitem{olnn}
R.~Narayanan and H.~Neuberger,
\NP{B412,1994,574};
\PRL{71,1993,3251};
\NP{B34{\rm (Proc. Suppl.)},1994,95}, 587;
\NP{B443,1995,305}.

\bibitem{twist}
R.~Narayanan and H.~Neuberger,
\PL{B348,1995,549}.

\bibitem{schwinger}
R.~Narayanan, H.~Neuberger and P.~Vranas,
\PL{B353,1995,507}.

\bibitem{notwave}
R.~Narayanan and H.~Neuberger,
\PL{B358,1995,303}.

\bibitem{majorana}
P.~Huet, R.~Narayanan and H.~Neuberger,
\PL{B380,1995,291}.

\bibitem{2dtorus_bc}
R.~Narayanan and H.~Neuberger,
UW-PT-96-04, Mar 1996.

\bibitem{shamir_anomaly}
Y. Shamir,
\NP{B417,1993,167}.

\bibitem{al}
S.~Aoki and R.B.~Levien,
\PR{D51,1995,3790}.

\bibitem{rs}
S.~Randjbar-Daemi and J.~Strathdee,
\PL{B348,1995,543};
\NP{B443,1995,386}.

\bibitem{eta_inv}
 D.B. Kaplan and M.~Schmaltz,
\PL{B368,1996,44}.

\bibitem{kaplan}
 D.B. Kaplan,
\PL{B288,1992,342};
\NP{B30{\rm (Proc. Suppl.)},1993,597}.

\bibitem{domainwall-ol-gs}
M.F.L. Golterman and Y. Shamir,
\PL{B353,1995,84}; Erratum-ibid. \andvol{B359,1995,422}.

\bibitem{waveguide}
M.F.L. Golterman, K. Jansen, D.N. Petcher and J. Vink,
\PR{D49,1994,1606};
M.F.L. Golterman and Y. Shamir,
\PR{D51,1995,3026}.

\bibitem{gribov}
T.~Aoyama and Y.~Kikukawa, in preparation.

\bibitem{gauge_2dXY}
W.~Bock, J.~Smit and J.C.~Vink,
\NP{B414,1994,73}.

\bibitem{cluster}
U.~Wolff, \PRL{62,1989,361}; \NP{B322,1989,759}; \NP{B334,1990,581}.

\bibitem{wz} A.M.~Polyakov and P.B.~Wiegmann, \PL{131B,1983,121};
\PL{141B,1984,223};
P.~Di Vecchia, B.~Durhuus and J.L.~Petersen, \PL{144B,1984,245}.

\bibitem{witt}E.~Witten, \CMP{92,1984,455}.

\bibitem{golt} M.F.L.~Golterman, K.~Jansen and D.B.~Kaplan,
\PL{B301,1993,219}.

\bibitem{kar} L.H.~Karsten and J.~Smit, \NP{B183,1981,103}.

\bibitem{kog} J.~Shigemitsu and J.B.~Kogut, \NP{B190[FS3],1981,365}.
\bibitem{gock} G.~Bahnot, A.~Gocksch and P.~Rossi, \PL{199B,1887,101};
A.~Gocksch, \PL{206B,1988,290}.

\bibitem{set-method}
G.~Bhanot, S.~Black, P.~Carter and R.~Salvador, \PL{183,1987,331}.

\bibitem{pure-gauge-model} H.~Hata, \PL{143B,1984,171}.
\bibitem{shamir_gaugefix_local} Y.~Shamir, TAUP-2306-95, Dec 1995.

\end{thebibliography}
\end{document}